%% file: libeskind.tex
\documentclass[12pt,usenatbib]{mn2e}
\usepackage{graphicx}
\usepackage{epsf}

\def\Mo{{\rm M_\odot}}
\begin{document}

\title[Satellite Galaxy Distribution]{The Distribution of Satellite
Galaxies: The Great Pancake} 
\author[N. I. Libeskind et al.]{
\parbox[h]{\textwidth}{Noam I Libeskind$^1$, Carlos S Frenk$^1$, Shaun
Cole$^1$, John C Helly$^1$, Adrian Jenkins$^1$, Julio F Navarro$^2$ 
and Chris Power$^3$}
\vspace{6pt} \\
$^1$Department of Physics, University of Durham, Science Laboratories,
South Road, Durham, DH1 3LE, U.K. \\
$^2$Department of Physics and Astronomy, University of Victoria, 
3800 Finnerty Road, Victoria, BC V8P 1A1, Canada \\
$^3$Centre for Astrophysics \& Supercomputing, Swinburne University of 
Technology, P.O. Box 218, Hawthorn, Victoria 3122, Australia \\
}
\maketitle


\begin{abstract} 
The 11 known satellite galaxies within 250~kpc of the Milky Way lie
close to a great circle on the sky. We use high resolution N-body
simulations of galactic dark matter halos to test if this remarkable
property can be understood within the context of the cold dark matter
cosmology.  We construct halo merger trees from the simulations and
use a semianalytic model to follow the formation of satellite
galaxies. We find that in all 6 of our simulations, the 11 brightest
satellites are indeed distributed along thin, disk-like structures
analogous to that traced by the Milky Way's satellites. This is in
sharp contrast to the overall distributions of dark matter in the halo
and of subhalos within it which, although triaxial, are not highly
aspherical. We find that the spatial distribution of satellites is
significantly different from that of the most massive subhalos but is
similar to that of the subset of subhalos that had the most massive
progenitors at earlier times. The elongated disk-like structure
delineated by the satellites has its long axis aligned with the major
axis of the dark matter halo. We interpret our results as reflecting
the preferential infall of satellites along the spines of a few
filaments of the cosmic web.

\end{abstract}


\section{Introduction}
\input{introduction.tex}

\section{Simulations and galaxy formation model}
\input{methods.tex}

\section{Results}
We begin by quantifying the shapes of dark matter halos in the
simulations and sub-sytems within them. We then interpret the results
in terms of the formation histories of the halos and their
subsystems. 
        \subsection{The morphology of halos and their subsystems}
	\input{results_1.tex}
	\subsection{Interpretation}
	\input{results_2.tex}

\section{Discussion and conclusions}
\input{conclusion.tex}

\end{document}

%% file: introduction.tex
\label{intro}

In the cold dark matter (CDM) cosmology, structure builds up 
through fragments merging together in a roughly hierarchical way. High
resolution N-body simulations of the formation of dark matter halos in
the $\Lambda$CDM cosmology have demonstrated that the cores of tightly
bound fragments often survive the merging process and remain as
distinct substructures orbiting inside a parent halo (\citealt{Klypin99},
\citealt{Moore99}). The centre of the main halo and the accompanying
substructures are naturally identified with the formation sites of
central and satellite galaxies respectively. The N-body simulations
suggest that the mass functions of surviving substructures in galactic
and cluster halos are roughly self-similar. Yet, the luminosity
function of galaxies in rich clusters has a very different shape from
the luminosity function of satellites in smaller systems such as the
Milky Way or the Local Group (\citealt{Kauffmann93}; \citealt{Mateo98}; \citealt{Trentham02}; \citealt{Benson02b}). Not only is the central galaxy much more
prominent in Milky-Way systems than in galaxy clusters, but the number
of surviving subhalos in simulations of galaxy-sized halos far exceeds
the number of the known satellites of the Milky Way.

The discrepancy between the small number of satellites around the Milky
Way and the large number of surviving substructures, once regarded
as a major challenge to the cold dark matter  cosmology, is now
thought to be due to the astrophysical processes that regulate the
cooling of gas in halos and its subsequent transformation into
stars. The increase in the entropy of the intergalactic medium brought
about by the reionization of the gas at early times has been
identified as a possible solution to the so-called ``satellite
problem'' (\citealt{Kauffmann93}; \citealt{bullock02}; \citealt{Benson02a}; 
see also \citealt{Stoehr02}). 
Reionization sharply reduces the efficiency of gas cooling in small
halos so that galaxies that formed prior to reionization are
preferentially those that end up as satellites in systems like the
Local Group. The detailed model calculated by
\cite{Benson02a} which includes the effects of early reionization as well as
other forms of feedback, reproduces many observed properties of the
Local Group's satellite system, including the distribution function of
circular velocity, the luminosity function and the colour
distribution.

While the original satellite problem is no longer deemed a serious
challenge, another related potential problem for the cosmological
paradigm has recently been highlighted by Kroupa, Thies \& Boily (2005).
 These authors
argue that the strongly flattened spatial distribution of the 11
brightest dwarf satellites of the Milky Way, a feature known, but not
understood, for many years (\citealt{LyndenBell82}; \citealt{Majewski94}), is
inconsistent with the $\Lambda$CDM model.  According to \cite{Kroupa05},
CDM models predict a roughly isotropic distribution of
satellites. They based this conclusion on the assumption that the
spatial distribution of satellites resembles the spatial distribution
of the halo dark matter which indeed, as N-body simulations have
demonstrated, is approximately (although not exactly) spherical (e.g.
\citealt{Frenk88}; \citealt{Jing02}; \citealt{bullock02}).

In this paper, we demonstrate that the satellites of systems like the
Local Group do {\it not} trace the distribution of halo mass. On the
contrary, the satellites in our suite of high resolution N-body
simulations are generally arranged in highly flattened configurations
which have similar properties to those of the Milky Way satellite
system. This, at first sight surprising, result is a reflection of the
anisotropic accrection of subhalos which generally stream into the
main halo along the filaments of the cosmic web. The flattened
structure in which the Milky Way satellites lie traces a great circle
on the sky and is approximately perpendicular to the Galactic
Plane. In our simulations, the satellite structures tend to be aligned
with the major axis of the triaxial halo mass distribution, that is,
the longest axis of the halo is close to lying in the principal plane
of the satellite distribution.

As this paper was nearing completion, two related papers appeared on
astro-ph.  Both of them used high-resolution simulations of galaxy
halos similar to those that we have performed. \cite{Kang05}
identified ``satellites'' in their 4 simulations with randomly chosen
dark matter particles taken either from the halo as a whole or
exclusively from substructures. They were able to find flattened
satellite systems similar to that of the Milky Way in the former case
but not in the latter. \cite{Zent05} found satellites in three
N-body simulations of Milky-Way type systems also in two different
ways. In the first, they used the semi-analytic model of
Kravtsov, Gnedin \& Klypin (2004) which is based on similar principles as those
applied by \cite{Benson02a}. In their second model, they 
identified satellites with the most massive subhalos.  \cite{Zent05}
found that in both cases, the satellite systems had a planar
distribution similar to that in the Milky Way and argued that the
degree of central concentration of the satellite systems plays an
important role in this result. They also showed that the population of
subhalos as a whole is anisotropic and preferentially aligned with the
major axis of the triaxial halo.

Like \cite{Zent05}, our study employs a semi-analytic model to follow
the formation of the visible satellites. In this respect, both these
studies are quite different from that of \cite{Kang05} who based their
conclusions purely on dark matter particles. Our model differs from
that of \cite{Zent05} in several important respects. Our simulation
codes and methods for identifying substructure are different. While
they considered three halos specifically chosen to lie on a filament,
we used 6 simulations randomly chosen from a large cosmological
volume. The biggest difference, however, concerns the semi-analytic
models used in the two studies. While both of them give a reasonable
match to several observed properties of the Milky Way's satellites,
our semi-analytic model has been applied and tested much more
extensively than that of \cite{KGK04}. The model we use is based
on the \texttt{GALFORM} code of \cite{Cole00} as extended by
\cite{Benson02b}. This model has been shown to give an acceptable
account of many properties of the galaxy population as a whole
including the luminosity function in various passbands, from the UV to
the far infrared, and in various environments, distributions of
colour, size and morphological type, etc. The model is also relatively
successful at matching the properties of galaxies at high-redshift, as
discussed in \cite{Baugh04}. Finally, the two studies use somewhat
different methods to quantify the distribution of Milky Way satellites
and to compare the results with the observations. On the whole, the
conclusions of the two studies are consistent although there remain
some differences as we discuss in Section~\ref{conclusion}.

The remainder of this paper is organised as follows: in
Section~\ref{Methods} we outline the methods used; in
Section~\ref{results} we present our results which we interpret in
Section~\ref{formhist}; in Section~\ref{conclusion} we discuss the
implications of our findings.

%% file: methods.tex
\label{Methods} We have analyzed 6 high-resolution $N$-body
simulations of galactic-size dark matter halos carried out with the
\texttt{GADGET} code (\citealt{Spring01a}). The halos, chosen to have a
mass $\sim 10^{12} \Mo$, were otherwise randomly selected from a large
cosmological simulation of a cubic region of side 35.325~$h^{-1}$ in a flat
$\Lambda$CDM universe (with $\Omega_{\rm m}=0.3, h=0.7,
\sigma_8=0.9$). The simulation was executed a second time adding
``high resolution'' (i.e. small mass)
particles, and appropriate small scale power in the initial
conditions, to a region surrounding the halo under
consideration. These simulations have been studied extensively in
previous papers (\citealt{Power03}, \citealt{Hayashi04},
\citealt{Navarro04}) and we refer the reader to those papers for specific
details of how the simulations were carried out. Table~\ref{simtable}
summarizes the important parameters of the simulations.

\begin{table}
\begin{center}
 \begin{tabular}{l l l l l l}
        & $N_{\rm tot}$ & $N_{\rm hr}$  &  $R_{\rm vir}$ & $N_{\rm vir}$\\
        & ($10^6$) & ($10^6$) &  ($h^{-1}$~kpc)  &  ($10^6$)\\
   \hline
   \hline
   gh1 & 14.6 & 12.9  & 110 & 1.07\\
   gh2 & 18.1 & 16.2  & 131 & 1.74\\
   gh3 & 18.0 & 16.2  & 170 & 3.73\\
   gh6 & 25.5 & 22.2  & 169 & 3.76\\
   gh7 & 19.2 & 17.3  & 156 & 2.99\\
   gh10& 13.4 & 12.1  & 133 & 1.86\\
    \hline
    \hline

 \end{tabular}
 \end{center}
\caption{Parameters for the six $N$-body halo simulations. For each
halo (column~1), (2) shows the total number of particles in the
simulation box, (3) the number of high resolution particles in
the simulation, (4) the virial radius of the halo in $h^{-1}$~kpc
defined as the distance from the centre at which the mean interior
density is $178\rho_{\rm crit}$, and (5) the number of particles
inside the virial radius. In all cases, the simulation cube has
co-moving length of 35.325 $h^{-1}$Mpc, and the mass of the high
resolution particles is $2.64\times 10^{5}h^{-1} M_{\odot}$ 
}
\label{simtable}
\end{table}

We identified bound substructures in the simulation using the
algorithm \texttt{SUBFIND} (\citealt{Spring01b}). First,
``friends-of-friends" groups (\citealt{DEFW85}) are found by linking
together particles whose separation is less than 0.2 times the mean
interparticle separation, corresponding roughly to particles within
the virialized region of the halo. \texttt{SUBFIND} then identifies
substructures within these halos based on an excursion set approach,
using the spatial and velocity information for each particle in order
to define self-bound objects.

For each halo, we generate a complete merger history, identifying all
progenitor and descendant halos, as described in
\cite{Helly03}. The semi-analytic galaxy formation model is calculated
along each branch of the merger tree. This is based on the model
described in detail in \cite{Cole00} and \cite{Benson02b}. The model
includes the following physical processes: (i) the shock-heating and
virialization of gas within the gravitational potential well of each
halo; (ii) radiative cooling of gas onto a galactic disk; (iii) the
formation of stars from the cooled gas; (iv) the effects of
photoionization on the thermal state and cooling properties of the
intergalactic medium; (v) reheating and expulsion of cooled gas
through feedback processes associated with stellar winds and
supernovae explosions (see \citealt{Benson03b}); (vi) the evolution of
the stellar populations; (vii) the effects of dust absorption and
radiation; (viii) the chemical evolution of the stars and gas; (ix)
galaxy mergers (which, depending on the violence of the merger, may be
accompanied by starbursts and the formation of a bulge -- see
\citealt{Baugh04}); (x) the evolution of the size of the disk and
bulge. 

Our model differs from that of \cite{Cole00} and \cite{Helly03} in the
way in which galaxy mergers are treated. In the current model, the
positions of satellite galaxies and the time when they merge is
determined by using information from \texttt{SUBFIND}.
 Central galaxies are placed
on the most bound particle of the most massive subgroup in the
halo. (\texttt{SUBFIND} identifies the background mass distribution of the halo
as a separate subgroup, so this is generally a robust way to define
the centre of the halo.) Satellite galaxies are placed on the
descendant subhalo of the progenitor halo in which they formed. If the
subhalo ceases to be identified by \texttt{SUBFIND} at some later output time,
we continue to trace its constituent particles and place the galaxy at
the centre of mass of this group of particles. A galaxy is considered
to have merged onto the central galaxy if its distance from the
central galaxy is less than the spatial extent of the set of particles
it is associated with.

An overview of the results of our semianalytic model as regards the
evolution of the galaxy population as a whole may be found in
\cite{Benson03a} and \cite{Baugh04} while results relevant to the
satellites of the Milky Way may be found in \cite{Benson02a}.

%% file: results_1.tex
\label{results}
\begin{figure}
\includegraphics[width=20pc]{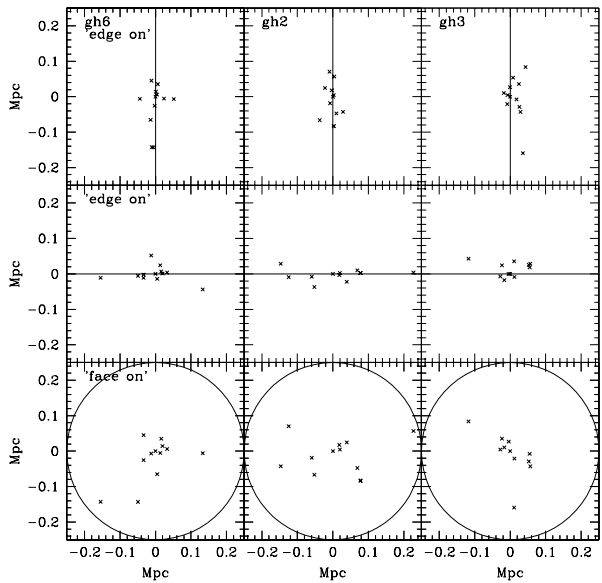}
\caption{Projections of the positions of the 11 most massive
satellites within 250~kpc of the central galaxy along the principal
axes of the inertia tensor in simulations gh6, gh2 and gh3.}
\label{sats1.3}
\end{figure}

The semianalytic model applied to the N-body simulations provides the
position and internal properties of the central galaxy in each halo
and its satellites. According to the semianalytic model, three of the
central galaxies are spirals and three ellipticals. For the purpose of
comparing with the analysis of \cite{Kroupa05}, we select the 11 most
massive satellites in each halo within a distance of 250~kpc from the
central galaxy. We calculate the moment of inertia tensor of this
satellite subsample, weighting each object equally, and obtain the
principal axes of the distributions.

Fig.~\ref{sats1.3} shows three orthogonal projections along the
principal axes of the satellite systems in three of our six
simulations.  The other three are very similar. The figure reveals,
remarkably, that the loci of the 11 most massive satellites define a
thin, disk-like structure around the central galaxy. As we show below,
in most cases, the satellite structure is aligned with the major axis
of its triaxial host dark matter halo.

\begin{figure}
\includegraphics[width=20pc]{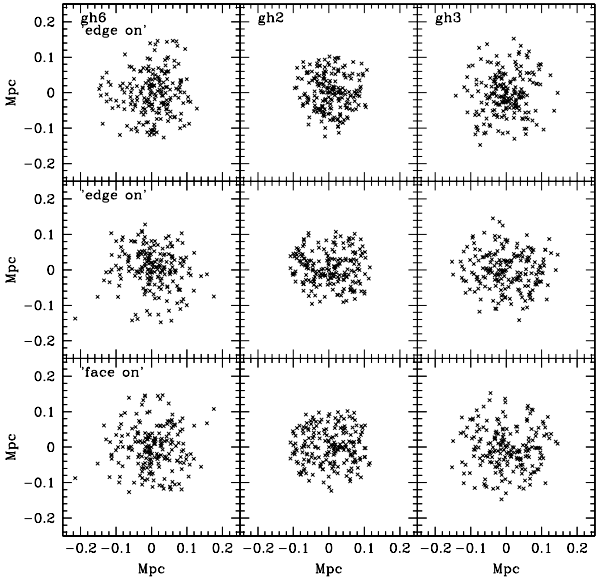}
\caption{Projections of the positions of the 200 most massive dark
matter substructures within 250~kpc of the central galaxy along the
principal axes of the inertia tensor in simulations gh6, gh2, and
gh3.}
\label{sg1.3}
\end{figure}

The distribution of the visible satellites differs significantly from
the distribution of the dark matter substructures identified by
\texttt{SUBFIND}. Fig.~\ref{sg1.3} is analogous to Fig.~\ref{sats1.3}
but the points plotted now correspond to the most massive 200
substructures found within 250~kpc of the central galaxy. The
projections are along the principal axes of the inertia tensor of the
substructure systems. It is evident that the distribution of
substructures is much less anisotropic than that of the satellites in
Fig.~\ref{sats1.3}.
\begin{figure}
\includegraphics[width=20pc]{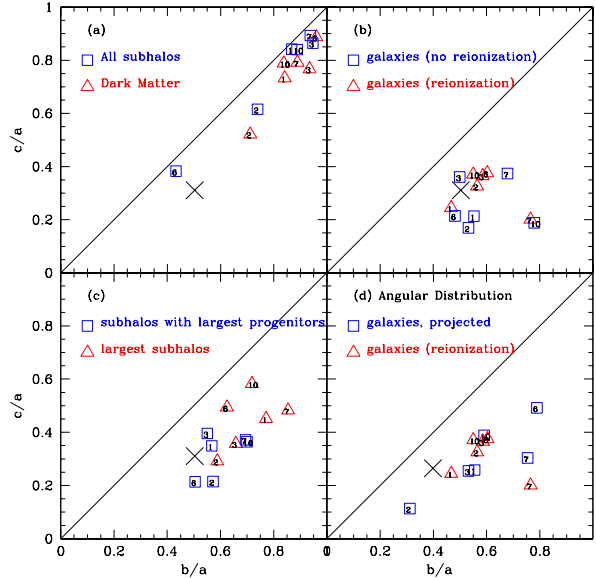}
\caption{Minor to major ($c/a$) versus intermediate to major ($b/a$)
axial ratios. Since $a>b>c$, the upper left triangular half of this
plot cannot contain any points. Along the diagonal lie prolate objects
and along the right vertical axis lie oblate objects. The numbers
inside each symbol identify the simulated halo. The 'X' indicates the
axial ratios of the Milky Way's satellite system from
\citealt{Kroupa05}. Only data out to a radius of 250~kpc is
used in all cases.  Panel
(a) compares the axial ratios of the dark matter halos (triangles)
with those of the system consisting of the dark
matter substructures (squares). Panel (b) compares the axial ratios of
the systems made up of the 11 most massive visible galaxies in models
with and without early reionization (triangles and squares
respectively). Panel (c) compares the axial
ratios of the systems of 11 most massive substructures with those of
the systems consisting of the 11 substructures that had the most
massive progenitors. Panel (d) compares the axial ratios of the systems made
up of the 11 most massive visible satellites in our full model
(triangles) with those of the same systems but with the radial
distances of each satellite normalized to a common value. This same
operation is peformed on the Milky Way data and so the
\citealt{Kroupa05} point moves slightly}
\label{caba} 
\end{figure}

The eigenvalues of the diagonalized inertia tensor are proportional to
the rms deviation of the $x$, $y$ and~$z$ coordinates relative to the
principal axes. Denoting the major, intermediate and minor axes by
$a$, $b$ and $c$ respectively ($a>b>c$), the flattening of the system
may be quantified by the ratios $c/a$ and $b/a$. The early N-body
simulations of \cite{Frenk88} showed that CDM halos are triaxial and
recent work indicates that c/a$= 0.7 \pm 0.17$, and b/a$> 0.7$
(\citealt{bullock02}).

The axial ratios, found by diagonalizing the moment of inertia tensor, 
of the dark matter halos and various subsystems of objects within them 
are plotted Fig.~\ref{caba}. Fig.~\ref{caba}a shows that
our simulated halos have axial ratios consistent with those found in
previous simulations and tend to congregate near the top right of the
panel corresponding to nearly spherical objects. With the exception of an 
outlier, gh6, which is significantly prolate, this is also the region
populated by the massive subhalos.

The axial ratios of the systems consisting of the 11 most massive
visible satellites are plotted in Fig.~\ref{caba}b. The triangles
correspond to our full semianalytic model and the squares to a
variant in which the early reionization of the intergalactic medium is
not included. The satellite systems in the two models have similar
flattening because more then 80 \% of the subhalos that host the brightest
satellites in the two cases are the same. 
However, as discussed by \cite{Benson02a}, neglecting the effects of
reionization leads to an overprediction of the number of faint
galaxies, including satellites in the Milky Way. Whether reionization
is included or not, the satellites in our simulations cluster around
the location of the Milky Way data marked by a cross in
Fig.~\ref{caba}b. This is the main result of our analysis: the
flattening of the satellite system in our simulations is in excellent
agreement with that of the Milky Way satellite system.

\begin{figure}
\includegraphics[width=20pc]{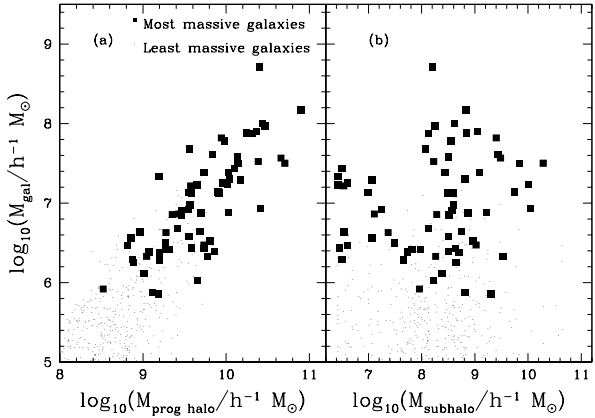}
\caption{(a) Stellar mass in each satellite galaxy as a function of the mass 
that its largest progenitor halo had before becoming incorporated into the
main halo. (b) Stellar mass in each satellite galaxy as a function of the mass 
of the substructure the galaxy currently resides in.
The squares show results for the 11 most massive satellites
in all 6 simulations, while the dots show results for the less
massive satellites. The overlap in panel (a) is largely due to the smallest 
satellite mass varying between simulations}
\label{mgmh}
\end{figure}

It is clear from Fig.~\ref{caba}a and~Fig.~\ref{caba}b that the most
massive visible satellites inhabit a biased subset of subhalos. To
explore the origin of this bias, we select two subsets of subhalos:
the 11 most massive subhalos at $z=0$ and the 11 subhalos which had
the most massive progenitors prior to being incorporated within the
virial radius of the main halo.  The flattening of these two systems
is compared in Fig.~\ref{caba}c. The figure shows that the crucial
factor in establishing a highly flattened system {\it is not} the
final mass of the subhalo but {\it the mass of the largest
progenitor}. It is the latter that correlates well with the final
stellar mass or luminosity of the visible satellite, as shown in
Fig.~\ref{mgmh}a. Here we plot the stellar mass of each satellite galaxy
against the mass of its largest progenitor. This strong correlation is a
result if the \texttt{GALFORM} model readily making the most luminous galaxies 
in the most massive progenitro halos. In contrast, Fig.~\ref{mgmh}b shows there
is no correlation between the stellar mass of each satellite and the mass
of its host substructure. This is due to the subhalos having been subjected 
to various amounts of tidal stripping.

Comparison of Fig.~\ref{caba}a and~Fig.~\ref{caba}c indicates that the flattening
of the systems consisting of the 11 most massive visible satellites
and the 11 subhalos that had the most massive progenitors are very
similar. This is an important result because it demonstrates that our
main conclusion regarding the compatibility of the \cite{Kroupa05}
data with the CDM cosmology does not depend on the details of our
semianalytic modelling of galaxy formation. So long as the brightest
satellites form in those subhalos with the most massive progenitors,
our conclusions stand.

With only 11 satellites in our main samples, the possibility that the
our estimate of the moments of inertia might be biased by the presence
of outliers is a concern. We investigate the sensitivity of our
results to outliers by scaling all radial positions to a common value
while keeping the angles of each radius vector fixed. The axial ratios
of the rescaled data are compared to the axial ratios of the 11 most
massive satellites in Fig.~\ref{caba}d. Rescaling the satellite
radial distances scatters our estimates of axial ratios somewhat but
does not, on average, lower the overall flattening of the
systems. As shown in the figure, rescaling the Milky Way data in the
same way also has a small effect on the axial ratios. 

Finally, we consider the connection between the highly anisotropic
distribution of satellites and the orientation of their host dark
matter halo. Consider the vector pointing along the major axis of the
distribution (i.e. along $a_{\rm sat}$).  Let $\theta$ denote the
angle between this vector and a vector pointing along the major axis
of the halo, $a_{\rm DM}$. For our six simulations, we find that
cos($\theta$) equals to 0.768, 0.979, 0.702, 0.747, 0.387,
0.942 for galaxy halos gh1, gh2, gh3, gh6, gh7 and gh10
respectively. Thus, apart from gh7, there is a very strong alignment
between the major axis of the disk-like satellite systems and the
major axis of the parent dark matter halo. In the Milky Way, the major
axis of the satellite disk-like structure is perpendicular to the
galactic disk. Thus, if our galaxy resides in a dark matter halo
similar to those that we have simulated, then the disk must be aligned
such that its normal vector points in the direction of the halo major
axis. Interestingly, this is exactly the alignment observed in recent
gasdynamical simulations of the formation of spiral galaxies by
\cite{NAS04}. 

%% file: results_2.tex
\label{formhist}
\begin{figure}
\begin{center}
\includegraphics[width=12pc]{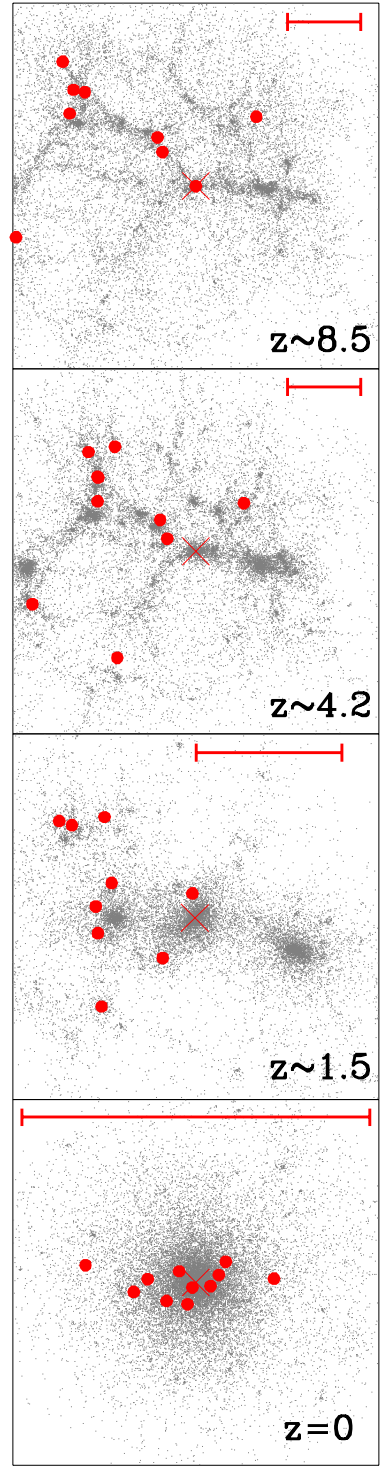}
\end{center}
\caption{The formation of a galactic halo and its satellites. The
points show a random 1\% of the dark matter particles that end up in the main
halo and the red circles the
positions of the 11 most massive satellites that end within 250~kpc of
the main galaxy by the present day. The scale of each plot is
indicated by the red line which has a co-moving length of
400 kpc. The initial collapse produces a 2D
structure -- a large pancake of dark matter.}
\label{sim1}
\end{figure}

The highly anisotropic distribution of satellite galaxies in Milky Way
type systems is a somewhat surprising outcome of galaxy formation in a
CDM universe. This is particularly so in view of the fact that the
population of subhalos as a whole is much less anisotropic and has
axial ratios similar to those of the halo dark matter. The key to
understanding the origin of the anisotropic satellite distribution
lies in the connection between halos and the cosmic web and, in
particular, in the way in which satellites are accreted onto the main
halo. Fig.~\ref{sim1} illustrates the anisotropic nature of satellite
accretion. The dots show a random 1\% of the dark matter particles that end up
in the main halo at the final time. The red circles mark the locations
of the most massive progenitors of the 11 satellites that have the
largest stellar mass at the final time.  Rather than originating
isotropically, those halos destined to become bright visible
satellites are accreted primarily along one or two of the cosmic web
filaments.

The first panel in the figure illustrates the highly anisotropic
collapse typical of CDM structures on galactic scales. Careful
inspection of the time evolution of the system shows that the collapse
occurs first along 2D sheet-like structures which subsequently wrap up
into filamentary streams of dark matter. By $z\sim 4.2$, these
filamentary ``highways'' along which proto-satellite galaxies form are
well established. The filaments are generally thicker than the locus of 
the largest proto-galactic halos which tend to
concentrate towards the central, densest parts of the filament in a
near 1-dimensional configuration. As the most massive halo progenitors
collapse to form the main galaxy, this alignment is largely
preserved. Smaller halos are more widely scattered across the thick
filaments, reflecting their weaker clustering strength
(\citealt{Cole89}; \citealt{MMW99}). In addition, they are often
accreted over a longer period and from a larger range of
directions. Their distribution, now lacking a preferred orientation,
ends up being much less anisotropic than that of the most massive
halos.  Whether reionization is included or not, satellite galaxies
in the semianalytic model form preferentially in the subhalos with the
most massive progenitors and thus inherit their highly flattened configuration.

%% file: conclusion.tex
\label{conclusion}

We have shown that the, at first sight surprising, flattened
distribution of satellites in the Milky Way is the natural outcome of
the anisotropic accretion of matter along a small number of filaments,
characteristic of halo formation in the CDM cosmology. \cite{Kroupa05}
reached the opposite conclusion, that the observed satellite
distribution is incompatible with the CDM model, because they
neglected the fact that the satellites do not trace the distribution
of halo dark matter but form instead in the most massive halos whose
spatial distribution is biased. 

Our results are not directly comparable to those of \cite{Kang05} who
also attempted to interpret the flattened distribution of Milky Way
satellites with the aid of high-resolution N-body
simulations. \cite{Kang05} assumed that the satellites follow the dark
matter distribution in the halo and did not consider the formation
sites of satellites in detail. \cite{Zent05}, on the other hand,
implemented a semi-analytic model similar to ours in N-body
simulations also similar to ours and those of \cite{Kang05}. 

Our results are broadly consistent with those of \cite{Zent05}. Unlike
them, we did not choose halos specifically lying along filaments but
selected them at random from a large cosmological simulation. In the
event, three of our halos would, according to our semianalytic model,
host spiral galaxies and the other three elliptical galaxies.  One
difference between the two studies is that
\cite{Zent05} found an acceptable match to the Milky Way satellite
distribution both in their semianalytic model and in a model in which
the satellites are identified with the most massive subhalos at the
final time. We have shown that the distribution of the latter is not
as flattened as the distribution of Milky Way satellites. The crucial
factor is not the final mass of the halo which is affected by tidal
stripping, but the mass of the largest progenitor before it is
accreted into the main halo. Indeed, if the satellites are identified
with the halos that had the largest progenitors, then their flattened
distribution is very similar to that of the satellites identified by
our semianalytic model. Thus, our conclusions are independent of the
details of our galaxy formation modelling.

As was also found by \cite{Zent05}, the major axis of the flattened
satellite distribution in our simulations points close to the
direction of the major axis of the parent halo. This alignment
reflects the preferential accretion of mass onto the halo along the
dominant filament.  An important consequence of this result is that if
the Milky Way resembles the systems we have simulated, then the
Galactic disk should lie in the plane perpendicular to the major axis
of the halo because the observed satellite system itself is
perpendicular to the Galactic disk. Interestingly, this is exactly the
configuration found in the recent gasdynamical simulations of spiral
galaxy formation by \cite{NAS04}.

The satellite alignment that we have found in our simulations is
almost certainly related to the ``Holmberg  effect,'' (\citealt{Holmberg69})the
observation that the satellites of external galaxies within a
projected radius of $r_{\rm p} \sim 50$~kpc tend to lie preferentially in a cone
along the galaxies' minor axis, avoiding the equatorial regions. To
test this observation requires a larger number of simulations than
those we have performed. Similarly, our current simulations are
inadequate to test the extension of the Holmberg effect uncovered by
\cite{ZSFW} from a study of isolated spirals which also revealed an
excess of satellites along the minor axis of the galaxy, now out to
projected distances of $r_{\rm p} \sim 500$~kpc.  A similar result was found by
\cite{Sales04} from a much larger sample 
of galaxies drawn from the 2 degree field galaxy redshift survey. They
too found an anisotropic distribution for $r_{\rm p} < 500$~kpc, but only
for satellites moving with a velocity relative to their host of
$\Delta v < 160$~km~s~$^{-1}$. In contrast,
\cite{Brainerd04} found the opposite effect in a sample of satellites from
the Sloan Digital Sky Survey, an alignment along the major axis at
small radii ($r_{\rm p} < 100$~kpc) and an isotropic distribution beyond.

Although our simulations are not large enough to study the
distribution of satellites beyond the inner 250~kpc of the galactic
centre, it seems likely that the anisotropic distribution of
satellites will continue out to larger separations. We intend to study
this problem in a larger set of simulations.

In summary, we have found that flattened distribution of the Milky Way
satellites, first noted by \cite{LyndenBell82}, and most recently highlighted 
by \cite{Kroupa05}, turns out to have a simple explanation in the context
of structure formation in the CDM model. It is merely a reflection of
the intimate connection between galactic dark matter halos and the
cosmic web.

%% file: libeskind.bbl
\begin{thebibliography}{99}
\bibitem[\protect\citeauthoryear{Baugh et al.}{2004}]{Baugh04}
Baugh C., Lacey C. G., Frenk C. S., Benson A. J., Cole S., Granato G. L., 
Silvia L., Bressan A., 2004, New A R, 48, 1239
\bibitem[\protect\citeauthoryear{Benson et al.}{2002a}]{Benson02a}
Benson A. J., Frenk C. S., Lacey C. G., Baugh C. M., Cole
S., 2002a, MNRAS 333, 177 
\bibitem[\protect\citeauthoryear{Benson et
al.}{2002b}]{Benson02b}Benson A. J., Lacey C. G., Baugh C. M.,
Cole S., Frenk C. S., 2002b, MNRAS, 333, 156
\bibitem[\protect\citeauthoryear{Benson et al.}{2003a}]{Benson03a}
Benson A. J., Bower R. G, Frenk C. S., Lacey C. G., Baugh C. M., Cole S.
2003a, ApJ, 599, 38 
\bibitem[\protect\citeauthoryear{Benson et al.}{2003b}]{Benson03b}
Benson A. J., Frenk C. S., Baugh C. M., Cole S., Lacey C. G., 2003b,
MNRAS, 343, 679 
\bibitem[\protect\citeauthoryear{Brainerd}{2004}]{Brainerd04} Brainerd
T., ApJL, in press (astro-ph/0408559) 
\bibitem[\protect\citeauthoryear{Bullock, Kravstov \& Weinberg} {2000}]{bullock00} Bullock J. S., Kravtsov A. V., Weinberg D. H., 2000, ApJ, 539, 517
\bibitem[\protect\citeauthoryear{Bullock} {2002}]{bullock02} Bullock, J. S. astro-ph/0106380
\bibitem[\protect\citeauthoryear{Cole \& Kaiser}{1989}]{Cole89} Cole S., Kaiser N., 1989, MNRAS, 237, 1127
\bibitem[\protect\citeauthoryear{Cole et al.}{2000}]{Cole00} Cole S., Lacey C. G., Baugh C. M., Frenk C. S., 2000, MNRAS, 319, 168
\bibitem[\protect\citeauthoryear{Combes}{2002}]{Combes02} Combes F., New A R, 46, 755 (astro-ph/0206126) 
\bibitem[\protect\citeauthoryear{Davis et al.}{1985}]{DEFW85} Davis M., Efstathious G., Frenk C. S., White S. D. M., 1985, ApJ, 292, 371
\bibitem[\protect\citeauthoryear{Frenk et al.}{1988}]{Frenk88} Frenk
C. S., White S. D. M., Davis M., Efstathiou G., 1988, ApJ, 327, 507 
\bibitem[\protect\citeauthoryear{Hayashi et al.}{2004}]{Hayashi04} Hayashi E., Navarro J. F., Power C., Jenkins A., Frenk C. S., White S. D. M., Springel V., Stadel J., Quinn T., 2004, MNRAS, 355, 794
\bibitem[\protect\citeauthoryear{Helly et al.}{2003}]{Helly03} Helly J., Cole S., Frenk C. S., Baugh C. M., Benson A., Lacey C. G., 2003, MNRAS, 338, 903
\bibitem[\protect\citeauthoryear{Holmberg}{1969}]{Holmberg69} Holmberg E., 1969, Ark. Astron., 5, 305 
\bibitem[\protect\citeauthoryear{Jing \& Suto}{2002}]{Jing02} Jing Y. P., Suto Y., 2002, ApJ, 574, 538
\bibitem[\protect\citeauthoryear{Kang et al.}{2005}]{Kang05} Kang X., Mao S., Gao L., Jing Y. P., A\&A submitted (astro-ph/0501333)
\bibitem[\protect\citeauthoryear{Kauffmann et al.}{1993}]{Kauffmann93} Kauffmann G., White S. D. M., Guiderdoni B., 1993, MNRAS, 264, 201 
\bibitem[\protect\citeauthoryear{Klypin et al.}{1999}]{Klypin99} Klypin A., Kravtsov A. V., Valenzuela O., Prada F., 1999, ApJ, 522, 82
\bibitem[\protect\citeauthoryear{Kravtsov et al.}{2004}]{KGK04} Kravtsov A. V., Gnedin O., Klypin A. A., 2004, ApJ, 609, 482  
\bibitem[\protect\citeauthoryear{Kroupa et al.}{2005}]{Kroupa05} Kroupa P., Thies C., Boily C. M., 2005, A\&A, 431, 517 (astro-ph/0410421)
\bibitem[\protect\citeauthoryear{Lynden-Bell}{1982}]{LyndenBell82} Lynden-Bell D., 1982, Obs., 102, 202
\bibitem[\protect\citeauthoryear{Majewski}{1994}]{Majewski94} Majewski S. R., 1994, ApJL, 431, L17
\bibitem[\protect\citeauthoryear{Mateo}{1998}]{Mateo98} Mateo M. L., 1998,
ARA\&A, 36, 435
\bibitem[\protect\citeauthoryear{Mo, Mao, \& White}{1999}]{MMW99} Mo
H. J., Mao S., White S. D. M, 1999, MNRAS, 304, 175 
\bibitem[\protect\citeauthoryear{Moore et al.}{1999}]{Moore99} Moore B., Ghigna S., Governato F., Lake G., Quinn T., Stadel J., Tozzi P., 1999, ApJ, 524, L19
\bibitem[\protect\citeauthoryear{Moore et al.}{2004}]{Moore04} Moore B., Kazantzidis S., Diemand J., Stadel J., 2004, MNRAS, 354, 522
\bibitem[\protect\citeauthoryear{Navarro et al.}{2004}]{Navarro04} Navarro J. F., Hayashi E., Power C., Jenkins A., Frenk C. S., White S. D. M., Springel V., Stadel J., Quinn T., 2004, MNRAS, 349, 1039
\bibitem[\protect\citeauthoryear{Navarro, Abadi \& Steinmetz}{2004}]{NAS04} Navarro J. F., Abadi M. G., Steinmetz M., 2004, ApJL, 613, L41
\bibitem[\protect\citeauthoryear{Power et al.}{2003}]{Power03} Power C.,
Navarro J. F., Frenk C. S., Jenkins A., White S. D. M., Springel
V., Stadel J., Quinn T., 2003, MNRAS, 338, 14 
\bibitem[\protect\citeauthoryear{Sales \& Lambas}{2004}]{Sales04} Sales L., Lambas D., 2004, MNRAS, 348, 1236 
\bibitem[\protect\citeauthoryear{Springel et al.}{2001a}]{Spring01a} Springel V., White S. D. M., Tormen G., Kauffmann G., 2001b, MNRAS, 328, 726
\bibitem[\protect\citeauthoryear{Springel et al.}{2001b}]{Spring01b} Springel V., Yoshida N., White S. D. M., 2001a, New Astronomy, 6, 51
\bibitem[\protect\citeauthoryear{Stoehr et al.}{2002}]{Stoehr02} Stoehr F., White S. D. M., Tormen G., Springel V., 2002, MNRAS, 335, L84
\bibitem[\protect\citeauthoryear{Susa \& Umemura}{2004}]{Susa04} Susa H., Umemura M., 2004, ApJ, 600, 1
\bibitem[\protect\citeauthoryear{Trentham \&
Hodgkin}{2002}]{Trentham02} Trentham N., Hodgkin S., 2002, MNRAS, 333, 423
\bibitem[\protect\citeauthoryear{White \& Rees}{1978}]{WR78} White
S. D. M., Rees M. J., 1978, MNRAS, 183, 341 
\bibitem[\protect\citeauthoryear{Zaritsky et al.}{1997}]{ZSFW}
Zaritsky D., Smith R., Frenk C., White S. D. M., 1997, ApJ, 478, L53 
\bibitem[\protect\citeauthoryear{Zentner et al.}{2005}]{Zent05} Zentner A. R., Kravstov A. V., Gnedin O. Y., Klypin A. A., ApJ submitted (astro-ph/0502496)

\end{thebibliography}
